\documentclass{aastex}

\usepackage{graphicx}
\usepackage{times}
\usepackage{emulateapj5}

\makeatletter

\newenvironment{inlinefigure}{%
\def\@captype{figure}%
\noindent\begin{minipage}{0.999\linewidth}\begin{center}}
{\end{center}\end{minipage}\smallskip}
\makeatother

\usepackage{amsmath}
\usepackage{apjfonts}

\def\arcmin{$^\prime$}
\def\arcsec{$^{\prime\prime}$}
\def\ergssec   {~ergs~s$^{-1}$}
\def\deg{$^{\circ}$}
\def\msun     {{M$_{\odot}$}}

\bibpunct{(}{)}{,}{a}{}{}

\shorttitle{NGC 4636}
\shortauthors{Jones et al.}

\begin{document}

\title{{\em Chandra} Observations of NGC 4636 - an elliptical galaxy in
turmoil}
\author{C. Jones, W. Forman, A. Vikhlinin, M. Markevitch, L. David,
A. Warmflash, S. Murray}
\affil{Harvard/Smithsonian Center for Astrophysics, 60 Garden St., MS-31, Cambridge, MA 02138}
\author{P. E. J. Nulsen}
\affil{Engineering Physics, University of Wollongong, Wollongong NSW
2522, Australia}

\begin{abstract}

Chandra images show symmetric, 8 kpc long, arm-like features in the
X-ray halo surrounding NGC~4636.  The leading edges of these features
are sharp and are accompanied by temperature increases of
$\sim30$\%. These properties, along with their scale and symmetry,
suggest that the arm-like structures are produced by nuclear
outburst driven shocks. We interpret these observations as part of a
cycle in which cooling gas originally fueled a nuclear
outburst about $3\times10^{6}$ years ago leading to shocks that reheat
the cooling gas, thus preventing the accumulation of significant
amounts of cooled gas in the galaxy center and temporarily starving
the central AGN.  

\end{abstract}


\section{Introduction}

NGC~4636 is one of the nearest and, at $L_X \sim$2$\times 10^{41}$
\ergssec, one of the most X-ray luminous ``normal'' ellipticals.
NGC~4636 lies in the outskirts of the Virgo cluster, 10\deg ~or 2.6
Mpc on the sky to the south of M87, for a distance to NGC~4636 of 15
Mpc (Tonry et al. 2001).  As found for most luminous, slowly rotating
ellipticals, the optical surface brightness of NGC~4636 flattens in the inner
regions.  The central region, as seen in short exposures and with HST,
has low eccentricity and is classed as an E0.  However, as noted by
Sandage (1961), at low surface brightness
the galaxy is flattened (E4).  NGC 4636 has
ionized gas in its core, but is unusual in that the gas velocities
are uncorrelated with and significantly larger than those of the
stars (Caon, Macchetto \& Pastoriza, 2000).  A weak, extended radio
source ($1.4 \times 10^{38}$ \ergssec) is observed at the galaxy
center (Birkinshaw \& Davies 1985, Stanger \& Warwick 1986), while
Loewenstein et al. (2001) place an upper limit of $2.7 \times 10^{38}$
\ergssec on nuclear X-ray emission.

Einstein X-ray images first showed that, like other luminous
elliptical galaxies, NGC~4636 was surrounded by an extensive
hot gas corona (Forman, Jones, \& Tucker 1985).  While Einstein observations
allowed only an emission-weighted temperature to be determined, with ROSAT
and ASCA, a modest increase in the gas
temperature with radius was found, abundances in the halo were measured and 
a very extended X-ray component was detected (Awaki et al. 1994, Trinchieri et
al. 1994, Matsushita et al. 1997, Finoguenov \& Jones 2000, Buote
2000).  From an Einstein HRI observation, Stanger \&
Warwick (1986) found an asymmetric gas distribution that they
suggested could be the result of erratic large scale gas flows.

The Chandra observatory allows us to study the structure of the X-ray
halo around NGC~4636 with a limiting resolution of $\sim$50 pc.  In
this paper we present an analysis of ACIS observations that reveal
unusual structures in the X-ray halo and describe a possible mechanism
for producing these features.

\section{Chandra X-ray Analysis}

NGC~4636 was observed with ACIS-S for 53 ksec on 26-27 January 2000
(obsid 323) and with ACIS-I for 11 ksec on 4-5 December, 1999 (obsid
324).  We filtered the observations by selecting ASCA grades 0, 2, 3,
4 and 6 and eliminating intervals of high background (``flares'') as
well as bright columns or pixels due to instrumental effects or cosmic
ray afterglows.  The remaining ``good'' times are 41286 seconds in the
ACIS-S observation and 5989 seconds in the ACIS-I observation.

Fig.~1 shows the region of the ACIS-S3 CCD centered on NGC4636 in the
energy range from 0.5 to 2.0 keV. This image shows a bright central
region surrounded by armlike structures.  At a distance of 15 Mpc,
these structures are observed to extend $\sim$8 kpc from the galaxy
center.  While the features toward the northeast and southwest are the
brightest and appear symmetric around the galaxy center, the ACIS-S
image shows a third, fainter arm northwest of the galaxy, that shares
symmetries with the two brighter X-ray arms.  In particular the sharp
edges along the bright southwest and faint northwest arms define a
parabola that also traces the southeast edge of the bright galaxy
core. Part of this parabola is mirrored in the bright NE arm.  In
addition to these features, Fig.~1 shows structure in the eastern part
of the halo.  In particular east of the galaxy center is a low surface
brightness region, with 1\arcmin\ elongated regions of enhanced
emission marking its northeast and southeast boundaries. On smaller
scales, as Fig.~2 shows, the X-ray emission in the galaxy core is
elliptical (position angle $\sim$320\deg) and has a 10\arcsec\ long
by 1\arcsec\ wide ``gap'' running north-south, where the surface
brightness is about half that in the adjacent regions.

\begin{figure*}[t]
  \centerline{\includegraphics[angle=-90, width=0.47\linewidth]{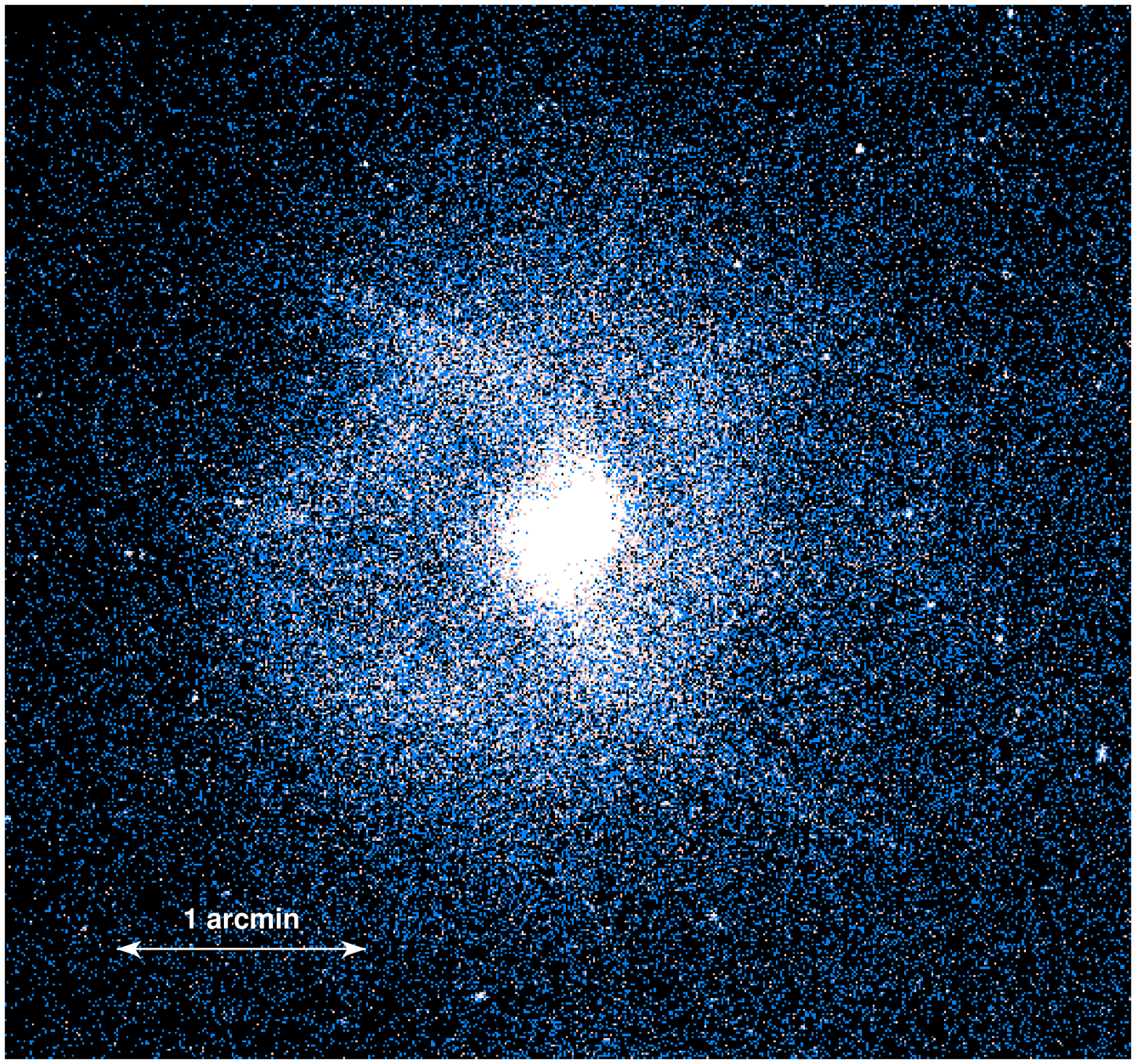}
  \includegraphics[angle=-90, width=0.455\linewidth]{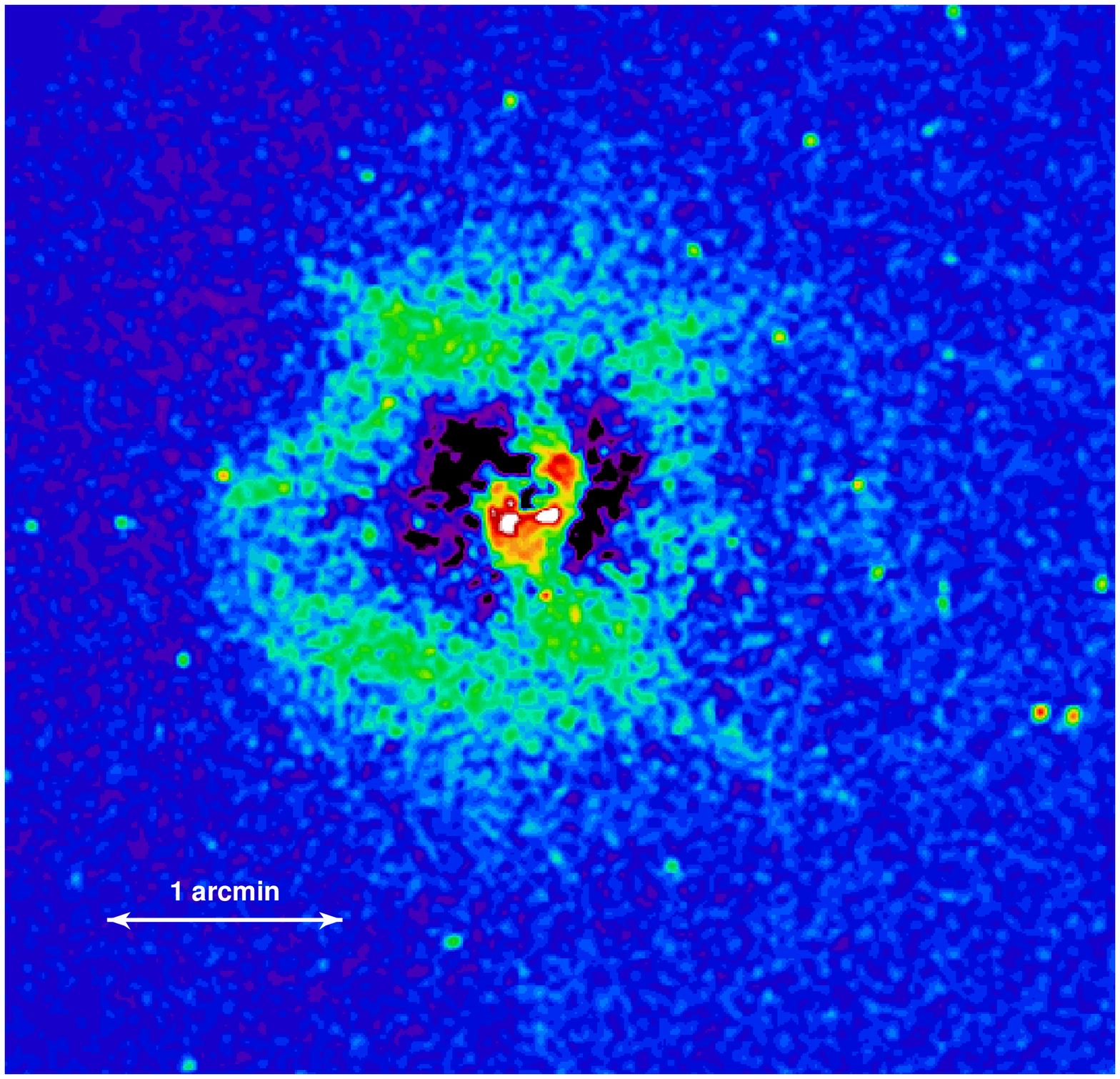}}

  \caption{The left panel shows the ACIS-S image of NGC~4636 in the 0.5
  to 2.0 keV energy at full resolution (1 pixel = 0.492\arcsec). We suggest
  shocks from a nuclear outburst produce  the brighter arm-like structures.
  The additional features could arise from another outburst. The right
  panel shows the emission after an azimuthally symmetric model
  describing the galaxy halo has
  been subtracted.  The remaining emission was smoothed with a two pixel 
  Gaussian.}
  \label{fig:panels}
  
\end{figure*}

To better define the halo structures, we generated a radial profile of
the X-ray emission excluding the brighter regions noted above, as well
as point sources, and used this profile to produce a smooth, radially
symmetric, nonparametric two dimensional model for the surface
brightness.  Fig.~1 (right) shows the result of subtracting this model
from the image on the left and smoothing the remaining
emission.  Most prominent in this figure are the enhanced regions of
emission and the cavities to the east and west of the nucleus.

Fig.~1 also shows extended X-ray emission that is brighter to the west of
the galaxy.  This is likely the very extended emission found from
ROSAT and ASCA observations (Trinchieri et al. 1994 and Matsushita et
al. 1997).  Although this extended emission was suggested to be symmetric
around NGC 4636, the extended emission in the Chandra ACIS-S image
(and in the ACIS-I image) is brighter west of the galaxy.

To study the bright armlike structures, we generated surface
brightness distributions across the NE and SW arms and measured the
gas temperature in several regions along and
across these arms. As an example, Fig.~3 (left) shows the X-ray surface
brightness distribution projected along a rectangle ($55'' \times
47''$) centered 65\arcsec from the nucleus and aligned parallel to the
``leading'' edge of the SW arm.  As suggested in Fig.~1, the
projections show that the ``leading'' edges are remarkably sharp.
Both arms show changes in brightness by about a factor of two on
scales of a few arcseconds.  The surface brightness
falls more gradually toward the trailing edge.

To measure the gas temperature in these features, and throughout
NGC~4636, we first identified and excluded regions around 127 point
sources in the ACIS-S3 CCD.  To identify sources we generated a 0.5 to
2.0 keV image without filtering for background flares. Chandra's
excellent angular resolution results in small ``cells'' for point
sources, particularly near the telescope aim point.  Thus for point
source detection, the benefit of the longer exposure time gained by
including times of high background outweighs the modest increase in
the local background. We omitted point sources with at least nine
counts.  We also examined a ``harder'' 2.0 - 5.0 keV image to insure
that no highly absorbed sources were missed.

We extracted spectra for a variety of regions from the ``cleaned
image'' and fit these using XSPEC over the energy range from 0.5 to 3
keV.  Although the focus of this paper is on the features in the
hot halo, we first comment on the overall temperature
structure.  For several regions within the central bright core, we
measured gas temperatures of 0.5 keV (with 90\% uncertainties less than
0.03 keV). Outside the central region, we observe a gradual
rise in temperature from 0.5 to 0.7 keV within a radius of three
arcminutes. This is consistent with the modest
rise in temperature with radius reported from ROSAT. 

For the NE and SW arms, we measured the gas temperature both along and
across these features.  Along the arms, we found the same increase in
gas temperature from 0.5 to 0.7 keV observed elsewhere in the halo.
To reduce any biases that this radial temperature change could produce
in looking for tempera-

\begin{inlinefigure}
  \centerline{\includegraphics[angle=-90, width=0.95\linewidth]{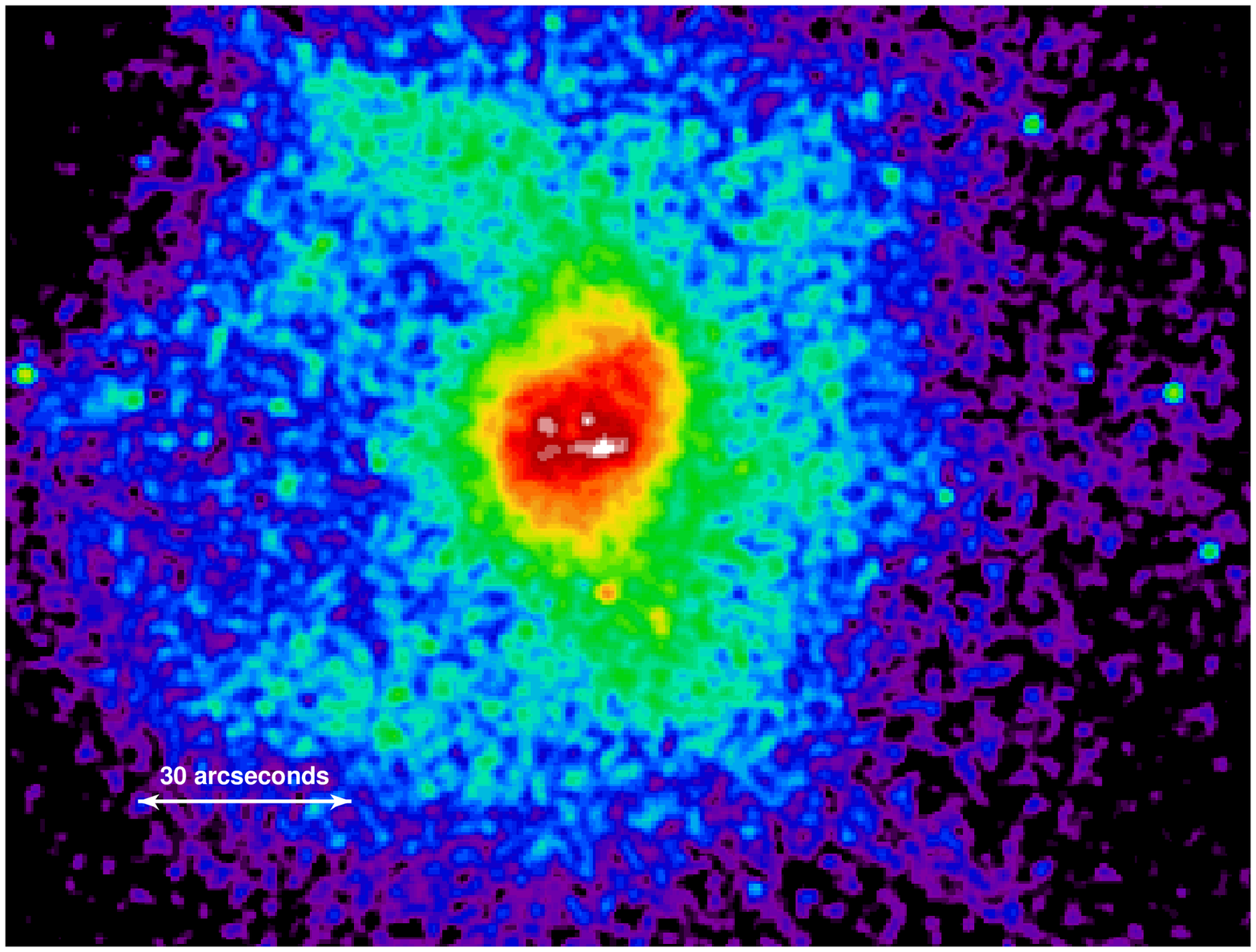}}
  \caption{The central region in the 0.5--2~keV
  band smoothed with a 1.5 pixel Gaussian shows structure in the core and the intersection of
  the NE arm with the northern part of the core and the SW arm with
  the southern part of the core.}
\label{fig:center}
\end{inlinefigure}

\noindent
ture changes across the arms, we extracted the
spectra across each arm in two regions, centered $50''$ and $80''$ from
the galaxy center.  For each region, we extracted spectra in a series
of rectangles ($4''$ by $27''$) whose long sides were parallel to the sharp
edge of each arm.  For the arm regions closer to the galaxy center,
where the halo emission is bright, we found no significant changes in
the gas temperature across the arms. In particular, in the inner
spectral regions of the NE arm, the best fit temperatures range from
0.57 to 0.65 keV (with 90\% uncertainties of 0.05 keV), while in the SW
inner arm, the best fit temperatures range from 0.60 to 0.68 keV.
However, in the outer region of the SW arm, the spectra show
significant changes in the gas temperature across the arm (see Fig.~3 right).
For this region of the SW arm, if we take the temperature measurement
south of the edge as the ambient temperature and fit the temperature
in the arm with two components, one fixed at the
temperature and normalization of the ambient region, the temperature
of the second component is $1.1\pm0.15$ keV. In the NE arm, the temperature
measured in the rectangle along the leading edge of the arm had the
highest temperature ($0.77^{+0.07}_{-0.13}$ keV), but this was not
significantly higher than in other regions.  The smaller and less
significant temperature changes in the NE arm may be due to
contamination from emission from the other structures in the eastern part
of the halo.

\section{A Nuclear Outburst in NGC 4636}

While the X-ray features in NGC 4636 are so far unique, they share
properties with structures seen in the hot gas in other galaxies and
in clusters.  The symmetric, parabolic regions of brighter emission
resemble the X-ray bright filaments observed around the radio lobes in
the elliptical galaxy M84 (Finoguenov and Jones 2000) and in the Hydra
A and Perseus clusters (McNamara et al. 2000, Fabian et al. 2000).
However in NGC~4636, no large radio lobes are observed. In addition
while the sharpness of the
edges of the NE and SW arms appears similar to the sharp edges
found along ``fronts'' in clusters (Markevitch et al. 2000,
Vikhlinin et al. 2001), the cluster ``fronts'' are
cold, while those in NGC~4636 are hot.

While the presence of sharp fronts suggests the possibility of
an ongoing merger, the east-west symmetry of the halo structures, the
similarity of this structure to that seen around radio lobes, as
well as the lack of a disturbed morphology in the stellar core 
or in the stellar velocities (Caon et al. 2000) led us
to consider a nuclear outburst as the underlying cause.
In particular we interpret the bright SW arm, the
fainter NW arm and the bright NE arm as the
projected edges of paraboloidal shock fronts expanding about an east
-- west axis through the nucleus.  A shock model is also consistent
with the evacuated cavities to the east and west of
the central region (Fig~1 right).

\begin{figure*}
  \centerline{\includegraphics[width=0.95\linewidth]{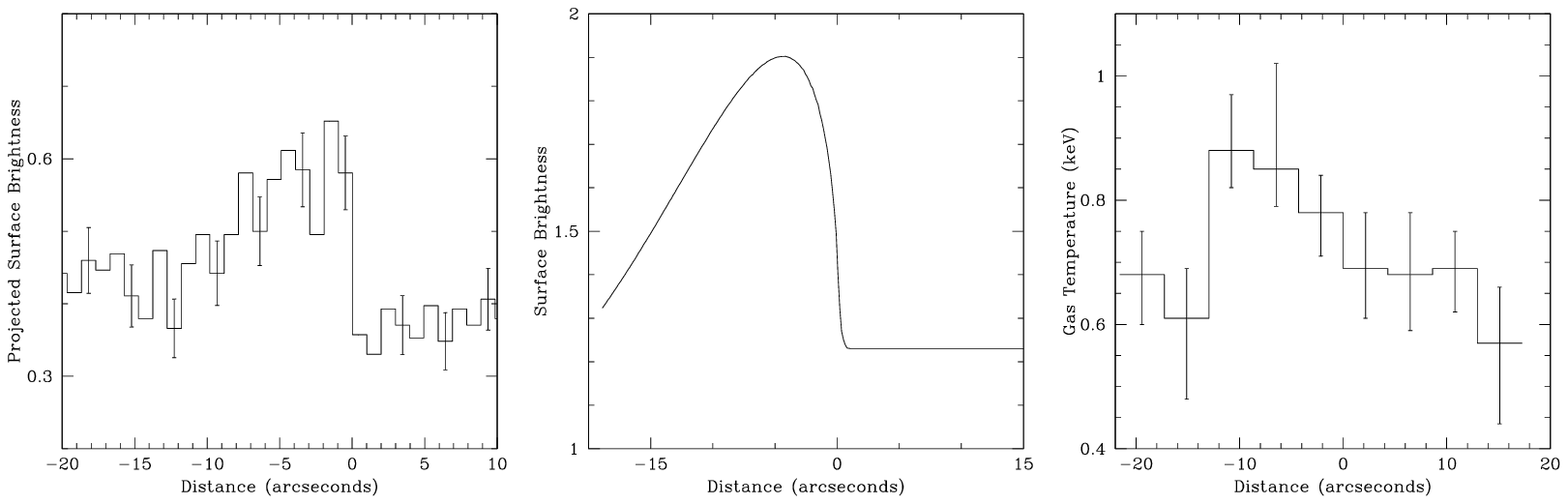}}
\caption{In each panel the surface brightness
discontinuity occurs at "0" on the x-axis. Left: Projected surface brightness across the SW arm at $\sim65''$ from the
galaxy center. Center: Expected surface brightness profile across the
shock front at $\sim65''$ from the galaxy center from the model.
Right: The projected temperature profile across the SW arm centered at
$\sim80''$ radius .  Error
bars are 90\% uncertainties.} 
\label{fig:profile}
\end{figure*}

With this interpretation, we used the size as well as the surface
brightness and temperature measurements of the SW arm to constrain a
simple shock model.  From our spectral measurements of the SW arm, the
gas outside the arm has an ambient temperature $kT \sim 0.65$ keV.
Taking the 1 keV temperature of the second spectral component in the
SW arm as the postshock temperature, we can estimate the strength of
the shock.  Such a temperature jump by a factor of 1.75 arises from a
shock having a density jump of a factor of 2. The corresponding
pressure jump is a factor of 3.5 and the Mach number is 1.73, giving a
shock velocity of 725 km s$^{-1}$.

We computed the evolution of a shock front produced by a point
explosion of $\sim 6 \times 10^{56}$ ergs in a uniform density gas.
The state of this model is completely determined by the strength of
the shock.  We chose a time when the density jump is a factor of two
for comparison with the observations.  In our model, the surface
brightness profile is determined by embedding the shocked region in a
uniform cube of unshocked gas.  The size of this cube relative to the
size of the shocked region determines the amount of emission from
unshocked gas, and hence the size of the jump in surface brightness
across the shock front.  The length of
the cube, $\ell$, is determined as the depth of gas, with the density
of 0.012 cm$^{-3}$ measured at $65''$ from the nucleus, that gives the
observed surface brightness at this radius.  Using beta model
parameters for the undisturbed gas measured from the X-ray surface
brightness profile ($\beta=0.45$, core radius $8''\!\!.4$) gives $\ell
= 144''$ (10.4 kpc).  At a radius of 65\arcsec, the diameter of the
shocked region in NGC 4636 is $\simeq107''$, so the ratio of $\ell$ to
the shock radius is $\sim 2.7$.  With this value, we obtain the
surface brightness distribution across the shock shown in Fig.~3. We
note that the shape of the shock front is not sensitive to the shock
strength.  The agreement between the surface brightness
profiles derived from the observations and from the model supports our
interpretation of these features as shock fronts. Closer to the
galaxy, where the preshock pressure is greater, the shock would
be weaker with a smaller temperature jump.  This may account for
the lack of an observed temperature jump in the inner arm regions.

Our analysis assumes that the symmetry axis of the shocks lies in the
plane of the sky. However, as shown in Fig~1 and 2, close to the
center of the galaxy, the bright rim of the western parabola can be
traced to the southeast of the nucleus, while the edge of the eastern
parabola can be traced to the northwest of the nucleus.  This
overlap of the bases of the parabolas (the shock fronts) implies that
the symmetry axis cannot be exactly in the plane of the sky.  To
determine the axis orientation, we would need
the intrinsic shape of the shocked regions.  This is
constrained by the agreement between the observed and model shock
profiles, and by the expectation that the shocked regions are not
highly elongated.  Both suggest that the symmetry axis is not
far from the plane of the sky, but detailed modelling is needed to
constrain this further.

Both the size and symmetry of the apparent shocks point immediately to
the nucleus as their energy source.  Scaling from our simple model,
the age of the shocks is $\sim 3\times10^6$ years and the total energy
driving them is $\sim 6 \times10^{56}$ ergs.  This implies a modest
mean nuclear output of $\sim 6\times10^{42}$ \ergssec.  However the
absence of either strong X-ray or radio emission suggests that the AGN
output is not constant over time.  If the power of the outburst were
near the Eddington limit for a $2\times10^8$ \msun black hole
(Magorrian et al. 1998), then the required energy could be released in
only $\sim1000$ years.  This raises the possibility of very brief
periods of quasar rebirth.  We suggest that the other emission features
in NGC~4636's  eastern halo originated from another nuclear
outburst. The east-west asymmetry may result from projection effects
or because the shocks have broken through to the east, but so far have been
confined by the denser medium west of the galaxy.
Such outbursts could cause the large chaotic velocities found in
the emission line gas.

The narrow waist in the shocked region close to the nucleus suggests,
in the simplest interpretation, that the energy driving the shocks was
injected off-center, possibly via jets.  Although the shape of the
shocked regions favors off-center injection, the presence of only a
weak, small-scale radio source in NGC 4636 argues against jets.
Alternatively, since the speed of a strong shock is $\propto
\rho^{-1/2}$, if the unshocked gas is sufficiently dense close to the
plane of the nucleus (perpendicular to the symmetry axis), then energy
injected close to the nucleus can be channeled to produce shocks of
this general shape.  In fact, there is no real distinction between
these; it is simply a matter of how far from the nucleus the energy is
injected.

\section{Discussion}

The hot gas in the centers of elliptical galaxies has a high density
and thus a short cooling time.  When the gas cools appreciably it
forms a dense disk due to rotation (Brighenti and Mathews 2000).  If
this cooled gas fuels a nuclear outburst that deposits energy close to
the nucleus, the presence of the dense disk would channel the energy
and produce shocks like those observed in NGC 4636.  Thus NGC 4636
appears to demonstrate the energy feedback process invoked to prevent
the deposition of large quantities of cooled gas in the centers of
galaxies and clusters (Tabor \& Binney 1993, Churazov et al. 2001,
David et al. 2001).  In a galaxy with the X-ray luminosity of
NGC~4636, outbursts would need to occur every $\sim 5\times10^7$ y to
prevent the accumulation of a significant amount of cooled gas.  If
outbursts causing shocks occur every $5\times10^7$ -- $5\times10^8$ y,
then NGC~4636 and other ellipticals could be in this shock phase 1 --
10\% of the time.  If the outburst is fueled by cooled gas, then the
repetition time must equal the central cooling time of the hot gas
following an outburst.  A typical elliptical galaxy, with a central
cooling time of a $\rm few \times 10^8$ years, would be about half way
through this cycle.  We recall that NGC~4636's X-ray luminosity is
unusually high for its absolute magnitude compared to other early type
galaxies, (Forman, Jones \& Tucker 1985).  This high X-ray luminosity
may be due to the advanced state of cooling that brought on the
outburst, and may be further enhanced by the shocks.  When the gas halo
returns to hydrostatic equilibrium, its X-ray luminosity could decline
by an order of magnitude or more, in which case the estimated outburst
intervals would be much less frequent.  Such excursions in X-ray
luminosity would help explain the broad $L_X$ -- $L_B$ relation for
elliptical galaxies.

Interactions between radio sources and the hot interstellar and
intergalactic medium in elliptical galaxies and clusters have been
observed in a number of systems, e.g. Cen A (Kraft et al. 1999), M84
(Finoguenov \& Jones 2000), Perseus (Bohringer et al. 1993), Hydra A
(McNamara et al. 2000), and A2052 (Blanton et al. 2001).
However, in these examples, the main observational characteristic
is displacement of hot gas by the radio emitting lobes, while in
approximate pressure equilibrium with the gas.  In NGC 4636 we see
evidence for a nuclear outburst, but no large radio lobes and only a
relatively weak (10$^{38}$ \ergssec) central radio source.
We suggest that the more common X-ray and radio lobe structures (e.g. M84)
represent a later stage of the process seen in NGC~4636.  This
raises the issue of whether the nuclear outbursts that provide
significant heating are primarily radio outbursts, or whether radio
outbursts occur in a phase following such an outburst. The latter scenario is
consistent with the modest ratio of jet mechanical power to radio
power found in Hydra A (McNamara et al. 2000). 

In summary, the Chandra image shows new features in the X-ray halo of
NGC~4636 that can be explained as the results of shocks produced by 
nuclear outbursts.  We suggest NGC 4636 is an example of the feedback
process that prevents the deposition of significant amounts of cold
gas by cooling flows.  The lack of extended radio jets or lobes
suggests that the energy from the AGN may be released by some more
direct process, rather than via a radio outburst.

\acknowledgements

We acknowledge stimulating discussions with E. Churazov, R. Mushotzky,
and H. Donnelly.  This work was supported by NASA contracts
NAS8-38248, NAS8-39073, the Chandra Science Center, and the
Smithsonian Institution.  PEJN gratefully acknowledges the hospitality
of the Center for Astrophysics.

\end{document}